\RequirePackage{fix-cm}
\documentclass[twocolumn]{svjour3}          % twocolumn
\smartqed  % flush right qed marks, e.g. at end of proof
\usepackage{graphicx}
\usepackage{amssymb}
\usepackage{xcolor}
\usepackage{bm}% bold math
\newcommand{\bvec}[1]{\boldsymbol {#1}}
\newcommand{\ket}[1]{| #1 \rangle}

\newcommand{\expected}[1]{ \langle #1 \rangle}

\journalname{Brazilian Journal of Physics}

\begin{document}

\title{Cold atoms beyond atomic physics}
%\titlerunning{Short form of title}

\author{Lucas Madeira$^1$ \and Vanderlei S. Bagnato$^{1,2}$}

\authorrunning{Lucas Madeira \and Vanderlei S. Bagnato} % if too long for running head

\institute{Lucas Madeira \at
\email{madeira@ifsc.usp.br} \\
$^1$
Instituto de F\'isica de S\~ao Carlos, Universidade de S\~ao Paulo, CP 369, S\~ao Carlos, S\~ao Paulo, 13560-970, Brazil. \\
$^2$
Hagler Fellow, Department of Biomedical Engineering, Texas A\&M University, College~Station,~TX~77843,~USA.
}

\date{Received: date / Accepted: date}

\maketitle

\begin{abstract}
In the last 25 years, much progress has been made producing and controlling Bose-Einstein condensates (BECs) and degenerate Fermi gases.
The advances in trapping, cooling and tuning the interparticle interactions in these cold atom systems lead to an unprecedented amount
of control that one can exert over them.
This work aims to show that knowledge acquired studying cold atom systems can be applied to other fields that share similarities and analogies with them, provided that the differences are also known and taken into account.
We focus on two specific fields, nuclear physics and statistical optics.
The nuclear physics discussion occurs with the BCS-BEC crossover in mind, in which we compare cold Fermi gases with nuclear and neutron matter and nuclei. We connect BECs and atom lasers through both systems' matter-wave character for the analogy with statistical optics.
Finally, we present some challenges that, if solved, would increase our understanding of cold atom systems and, thus, the related areas.

\keywords{Cold atoms \and Atomic physics \and Nuclear physics}
\end{abstract}

\section{Introduction}
\label{intro}

Much progress has been made producing and controlling Bose-Einstein condensates (BECs) \cite{Anderson1995,Bradley1995,Davis1995} and degenerate Fermi gases \cite{Ohara2002,Zwierlein2005} of dilute atomic clouds in the last 25 years \cite{Pethick2008}.
The advances in trapping, cooling, and tuning the interparticle interactions in these cold atom systems make them excellent candidates for studying
microscopic interactions
due to the amount of control that one can exert over these systems.
The physics learned in them could lead to progress in other systems and fields, as long as we can understand the similarities and take into account the differences between them.
In this work, we considered two fields that can be related to cold atoms: nuclear physics and statistical optics.

First, we start with nuclear physics.
The main reason nuclear and atomic physics share many concepts and techniques is because of the short-range character of the interactions.
If we take a characteristic length unit $\ell$ (for example, the typical interparticle distance) and the mass $m$, we can construct an energy scale $E_0=\hbar^2/(m\ell^2)$. Casting all distances in $\ell$ units and all energies in $E_0$ units allows us to focus on the \ scale-independent differences \cite{Zinner2013,Amorim1997,Riisager2000}.
An important quantity is the number density $n$ multiplied by the scattering length $a$ to the third power, $na^3$.
A mean-field approach yields good results when $na^3\ll 1$.
However, strong correlations appear when $na^3 \gtrsim 10^{-1}$ for atoms with $a/\ell\sim 100$ and nuclei with $a/\ell\sim 1$, provided they are away from resonance.

The focus of the comparison is centered around fermionic cold gases since protons and neutrons are fermions.
Moreover, we chose to discuss the similarities and differences 
under the light of the smooth connection between
Bardeen-Cooper-Schrieffer (BCS) superfluidity \cite{Bardeen1957} and Bose-Einstein condensation \cite{Bose1924},
the BCS-BEC crossover. As we will see, this formulation is quite convenient to compare both areas \cite{Zinner2013,Strinati2018}.
Special attention is devoted to neutron-neutron pairing since the parallel with two-component atomic Fermi gases can be readily established.

One of the goals of this review was to connect the cold atoms and nuclear physics communities. Hence, we tried to keep the discussion at a level interesting to readers from both backgrounds, and we provided references to more in-depth discussions of specific topics.

The second comparison is made with statistical optics.
Reference \cite{Tavares2017} found that
a ground-state BEC and a turbulent BEC \cite{Madeira2020,Madeira2020b} share an analogy with
the propagation of an optical Gaussian beam and elliptical speckle light map.
This occurs because both are examples of coherent matter-wave systems.
In principle, this could be used to increase our understanding of quantum turbulence by
looking at statistical atom optics.

This work is structured as follows.
Section~\ref{sec:nuclear} contains the comparison between cold atoms and nuclear physics.
First, we briefly discuss two-component atomic Fermi gases, Sec.~\ref{sec:twocomp}, since the concepts will be needed later to present the BCS-BEC crossover and discuss neutron matter.
In Sec.~\ref{sec:crossover_cold}, we introduce the BCS-BEC crossover and the unitary regime with cold atoms in mind.
The BCS-BEC crossover from the nuclear physics perspective is discussed in
Sec.~\ref{sec:bcsbec_nuclear}, including nuclear and neutron matter, alpha particle condensation, and nuclei.
In Sec.~\ref{sec:laser}, we show how BECs and statistical optics share some similarities.
Finally, in Sec.~\ref{sec:conc}, we list some challenges that permeate both cold atoms and the related areas.

\section{Nuclear physics}
\label{sec:nuclear}

\subsection{Two-component atomic Fermi gases}
\label{sec:twocomp}

The production of degenerate Fermi gases is routine in many experimental groups world-wide \cite{Bloch2008,Giorgini2008}. 
However, this was not the case in the early days of cold fermionic gases.
Evaporative cooling of a single species of fermions is not possible due to the Pauli exclusion principle. Hence, alternative techniques had to be developed \cite{Onofrio2016}.
If instead a mixture of two different fermionic species is prepared, then low-energy $s$-wave interactions between the species can be used to produce a degenerate gas \cite{Ketterle2008}.
The typical Fermi temperature is approximately 1 $\mu$K, while experiments reach hundreds of nK temperatures. The number density is close to $10^{13}$ cm$^{-3}$, which means that the typical interparticle spacing is close to 500 nm, much larger than the interatomic potential ranges (1-10 nm) and atom sizes (0.1 nm).

The quantum numbers related to the atoms are ($\bvec{J}$, $\bvec{I}$, $\bvec{F}$, $m_F$), $\bvec{J}$ being the addition of the orbital angular momentum $\bvec{L}$ and spin $\bvec{S}$ of the electrons, $\bvec{I}$ the total nuclear spin, and $\bvec{F}$ the total spin (addition of $\bvec{J}$ and $\bvec{I}$) with projection $m_F$. In the presence of a magnetic field, only $m_F$ is conserved, although good asymptotic quantum numbers exist in the limits of very low or high field strengths \cite{Pethick2008}.
The hyperfine energy splitting due to the magnetic field is typically of the order of $10^{-6}$ eV, which is much smaller than any possible electronic transition.
Two-component Fermi gases correspond to populations of the lowest two hyperfine states (two different values of $m_F$).
It is also worth noting that
three-component fermionic gases have been produced \cite{Ottenstein2008,Huckans2009,Nakajima2010}, which correspond to three $m_F$ values.

The atomic clouds are trapped by external potentials, which adds an energy scale to the system, the spacing between the trap eigenstates.
Let us consider, for simplicity, the case of an isotropic harmonic oscillator.
Typically, the energy $\hbar\omega$ is much smaller than the hyperfine splitting, sometimes by five orders of magnitude.
Hence, only two different internal states (corresponding to different $m_F$ values) can effectively describe a two-component gas of fermions.
After this discussion, we hope that it is clear why the theoretical oversimplification of treating a two-component Fermi gas as a mixture of spin-up and spin-down components produces the desired results.

\subsection{BCS-BEC crossover in cold atoms}
\label{sec:crossover_cold}

\begin{figure}[!htb]
\centering
  \includegraphics[width=\linewidth]{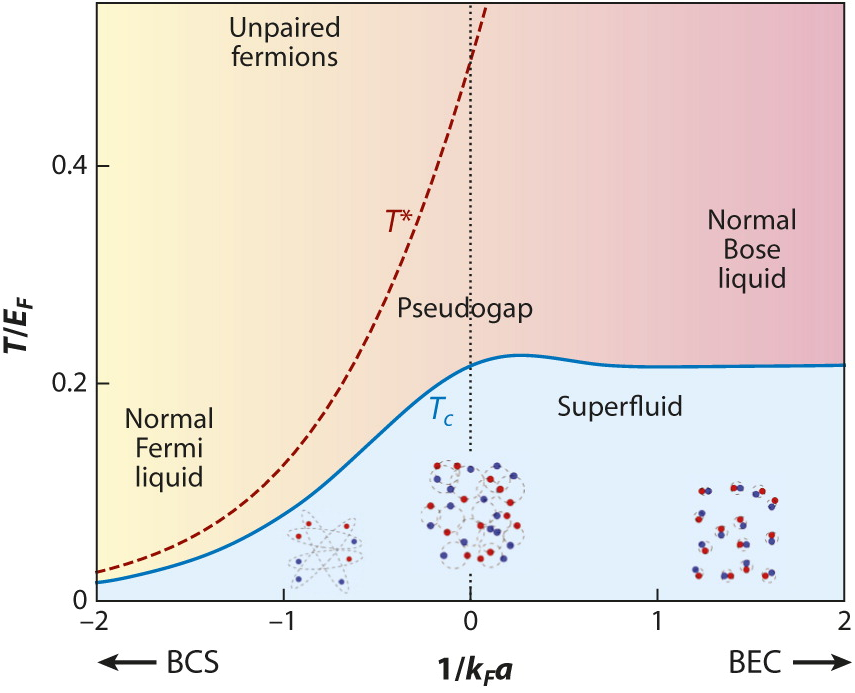}
\caption{
Phase diagram of the BCS-BEC crossover as a function of temperature $T/E_F$ and interaction strength $(1/k_F a)$.
The schematic evolution from the BCS limit with large Cooper pairs to the BEC limit
with tightly bound molecules is shown at the bottom. The unitary point, $1/(k_F a)= 0$, corresponds to strongly interacting pairs with size
comparable to $k_F^{-1}$. Taken from Ref.~\cite{Randeria2014}.}
\label{fig:1}
\end{figure}

Besides the one-body external field discussed in the previous section, two-body interactions between the atoms also introduce an energy and length scale into these systems.
Interparticle interactions in ultracold atomic gases can be tuned via Feshbach resonances, thus realizing the BCS-BEC crossover, a problem of significant interest \cite{Ketterle2008}. The investigation of Bardeen-Cooper-Schrieffer (BCS) to Bose-Einstein condensate (BEC) crossover arises in an attempt to better understand superfluidity and superconductivity beyond the standard paradigms \cite{Zwerger2011,Randeria2014}.
The crossover concepts can be traced back to considerations regarding strongly interacting superconductors by Eagles \cite{Eagles1969} and fermionic cold atoms by Leggett \cite{Leggett2008}.
The key idea is that the interparticle (electrons or alkali atoms) two-body interaction can be varied so that increasing its strength allows the system to go from a BCS paired state to a molecular BEC, where the two-body bound state is the relevant characteristic, Fig.~\ref{fig:1}.
The most exciting results are related to a very strongly interacting state of matter, the unitary Fermi gas, which is right at the crossover's heart. Until recently, all superconductors and superfluids fell into one of two classes, bosonic and fermionic. This led to two different paradigms, BEC and BCS, for understanding the properties of quantum fluids.

In the BEC scheme, first developed for noninteracting bosons and later generalized to include repulsive interactions, it is possible to describe bosonic fluids, such as $^4$He, and ultracold Bose gases, $^{87}$Rb for instance. The condensate is a macroscopic occupation of a single quantum state that occurs below a transition temperature $T_c$, which is of the same order of magnitude as the quantum degeneracy temperature, at which the interparticle spacing is of the order of the thermal de Broglie wavelength.

The BCS paradigm, first conceived for metallic superconductors, describes a pairing instability arising from a weak attractive interaction in a highly degenerate system of fermions. Both the formation of pairs and their condensation occur at the same temperature $T_c$, which is orders of magnitude smaller than $E_F/k_B$, which sets the degeneracy temperature. The BCS theory is very successful in describing conventional superconductors and it has been generalized to describe various systems, such as superfluidity in $^3$He and pairing in nuclei.

Early theoretical work on the crossover was conceptually interesting, but real enthusiasm came from its experimental realization \cite{Regal2004}.
The most important difference between ultracold Fermi gases and all previously studied superfluids is that the interaction between spin-up and spin-down can be tuned in the laboratory. The average separation between atoms, $k_F^{-1}$, is much larger than the range of the interatomic potential $r_0$. For a dilute gas with $k_F r_0 \ll 1$, the interaction is described by a single parameter, the s-wave scattering length $a$. All thermodynamic properties of the gas can be written in a universal scaling form. For example, the free-energy $F$ takes the form
\begin{equation}
F = N E_F \mathcal{F}(k_B T/E_F,1/(k_F a)),
\end{equation}
where $\mathcal{F}$ is a scaling function. This result is universal in the sense that it is independent of microscopic details, as long as $k_F r_0 \rightarrow 0$.

A Feshbach resonance is a dramatic increase in the collision cross-section of two atoms due to a bound state in the closed channel crossing the open channel's scattering continuum.
Let us consider the specific example of $^6$Li, of electronic spin $S=1/2$ and nuclear spin $I=1$. The electric spin is fully polarized, usually the magnetic field $B \geqslant 500 \ G$, and aligned in the same direction for each of the three lowest hyperfine states. Hence, two colliding $^6$Li atoms are in a continuum spin-triplet state in the open channel. The closed channel has a singlet-bound state that can resonantly mix with the open channel due to the hyperfine interaction that couples the electron spin to the nuclear spin.

The difference in the magnetic moments in the closed and open channels allows experimentalists to use an external magnetic field $B$ to tune a Feshbach resonance. The resulting interatomic interaction in the open channel can be described by a $B$-dependent scattering length which, near the resonance, is \cite{Moerdijk1995}
\begin{equation}
a(B) = a_{BG} \left[ 1 - \frac{\Delta B}{B-B_0} \right],
\end{equation}
where $a_{BG}$ is the background value, in the absence of the coupling to the closed channel, $B_0$ is the location of the resonance, and $|\Delta B|$ is the width.

Let us consider the problem of two fermions with spin-up and spin-down interacting with a two-body potential with range $r_0$. The low-energy properties as a function of the momentum $k$, such that $k r_0 \ll 1$, are described by the s-wave scattering amplitude
\begin{equation}
f(k)= \frac{1}{k \cot (\delta_0(k))-ik} \approx -\frac{1}{1/a + ik}.
\end{equation}
The scattering length $a$ completely determines $\delta_0(k \to 0)=-\tan^{-1}(ka)$, the s-wave scattering phase shift with low-energy. The effective interaction is independent of the shape of the potential, thus we can choose the simplest one: a square well of depth $V_0$ and range $r_0$. 

If we start with a very shallow square-well,
$a<0$ for weak attraction,
the scattering length
grows in magnitude with increasing $V_0$, and diverges to $-\infty$ at the formation of a bound state in vacuum ($V_0=\hbar^2 \pi^2/(m r_0^2)$, where the reduced mass is $m/2$). Once the bound state has been formed, the scattering length changes to $a>0$ and decreases from $+ \infty$ with increasing $V_0$. For positive $a$, the scattering length $a$ is the bound state's size, with energy $-\hbar^2/(ma^2)$.

The threshold for bound-state formation, where $|a|\rightarrow \infty$, is called unitary point. The phase shift is $\delta_0(k=0)=\pi/2$, and the scattering amplitude takes the maximum value $f \approx -1/(ik)$. The unitary regime is the most strongly interacting regime in the BCS-BEC crossover.
The unitary point is of great interest since that $|a|\to \infty$ means that the system loses a length scale, and the scattering length cannot be used as an expansion parameter, as noted by Bertsch\footnote{The challenge proposed by Bertsch to the participants of the Tenth International Conference on Recent Progress in Many-Body Theories
can be stated as:
what are the ground state properties of the many-body system composed
of spin-1/2 fermions interacting via a zero-range, infinite scattering
length contact interaction?
} \cite{Baker1999}.
For a fermionic system, it is expected that the energy is proportional to the only scale in the problem, the
non-interacting Fermi gas energy $E_{FG}$,
\begin{equation}
E_0=\xi E_{FG} = \xi \frac{3}{10} \frac{\hbar^2 k_F^2}{m},
\end{equation} 
where the constant $\xi$ is known as the Bertsch parameter.
In the limit $a k_F\rightarrow - \infty$,
quantum Monte Carlo (QMC) results give the exact value of $\xi=0.372(5)$
\cite{Carlson2011}, in agreement with experiments \cite{Ku2012,Zurn2013}.
 
Quantum Monte Carlo methods have been successfully employed to study several aspects of the BCS-BEC crossover besides the Bertsch parameter.
The term QMC broadly encompasses several methods, but in the cold atoms context, the most important ones are the variational and diffusion Monte Carlo \cite{Foulkes2001} and auxiliary-field QMC \cite{Carlson2011}.
Their goal is to provide an accurate solution, or approximation with controllable errors, to the many-body Schr\"odinger equation.
We should point out that these methods are also commonly employed in nuclear physics \cite{Carlson2012,Carlson2015,Lynn2019}.

Since the seminal work with the unitary Fermi gas \cite{Carlson2003} and
properties calculated across
the BCS-BEC \\ crossover \cite{Astrakharchik2004,Chang2004}, QMC methods have been applied to investigate several other aspects of two-component Fermi gases \cite{Gandolfi2011b,Bulgac2012,Carlson2013,Gandolfi2014}.
A few examples include:
effects of the effective range of the interaction \cite{Forbes2011,Forbes2012},
truly zero-range calculations \cite{Pessoa2015,Pessoa2015b,Pessoa2019},
the contact parameter \cite{Hoinka2013},
two-dimensional systems \cite{Galea2016,Galea2017},
the excitation of vortices in 3D \cite{Madeira2016,Madeira2019} and 2D \cite{Madeira2017} systems, and
mass imbalance systems, for example $^6$Li-$^{40}$K, between the two species \cite{Gezerlis2009}.

\subsection{The BCS-BEC crossover in nuclear physics}
\label{sec:bcsbec_nuclear}

Nucleons (protons and neutrons) are spin-1/2 particles, hence nuclear systems deal with four kinds of fermions: spin-up and spin-down protons and neutrons
($p\uparrow$, $p\downarrow$, $n\uparrow$, and $n\downarrow$).
The interactions between
two protons ($pp$), two neutrons ($nn$), or a proton and a neutron ($pn$) are nearly
identical. A convenient formulation that takes advantage of this
charge independence consists of adding a degree of freedom to each
nucleon called isospin.
If we consider protons
to be the analogous state of a spin-up particle, and neutrons
to be the counterpart of spin-down particles, we can benefit from
the algebra already developed for spin-1/2.
Protons and neutrons are isospin-1/2 particles, and a common
basis choice is $\bvec{T}_z\ket{p}=+(1/2)\ket{p}$ and $\bvec{T}_z\ket{n}=-(1/2)\ket{n}$.
Our discussion of the BCS-BEC crossover in nuclear physics follows Ref.~\cite{Strinati2018} closely, where the reader can also find some topics not covered here, such as finite temperature results.

The only two-nucleon bound state is the deuteron, which corresponds to the
$pn$ channel. With Sec.~\ref{sec:crossover_cold} in mind, we can imagine a crossover from a BEC of deuterons to a BCS state of $pn$ Cooper pairs.
The interaction between nucleons is fixed, contrary to cold atoms, where the scattering length can be tuned with great precision. However, the nucleon systems' density can be varied, thus changing the value of $k_F a$.
For example, at low densities, the deuteron is bound. Increasing the density decreases the binding energy to a point where the nucleus turns into a $pn$ Cooper pair.

In nuclei, which would be best described as ``clusters'' or ``droplets'' in the cold atoms context, $nn$ and $pp$ pairing are favored over $pn$ simply because there are usually more neutrons than protons. Although there is no equivalent of the BEC regime, due to $pp$ and $nn$ being unbound, the $nn$ scattering length is much larger than the interparticle spacing and effective range, thus leading to a scenario similar to the unitary limit.

Another possibility for a BCS-BEC crossover in nuclear physics involves the $\alpha$ particle, the bound state of two protons and two neutrons.
Its binding energy is much larger than the deuteron one, suggesting that a BEC of $\alpha$ particles would be energetically more favorable than a deuteron BEC. However, the $\alpha$ particle is much more sensitive to finite density effects. Increasing the density makes the $\alpha$ particle BEC turn into a BCS state of $pn$, or $pp$ and $nn$, Cooper pairs.

In the following sections, we explore each of the possibilities above in detail.

\subsubsection{The deuteron and symmetric nuclear matter}
\label{sec:sym}

The only bound state of two nucleons is the deuteron.
The $pn$ that compose the deuteron are in a spin-triplet state ($S=1$), isospin singlet ($T=0$), and total angular momentum $J=1$. It has angular momentum components $L=0$ and 2, the latter being due to the non-central tensor force.
It has only one bound-state, all of 
its excited states are scattering states.
Another evidence that the deuteron is very weakly bound is that
its binding energy is approximately 2.23 MeV. That is only
0.1\% of its rest mass, while heavier nuclei have binding energies of
$\approx 0.8$\% of their rest masses.
Another characteristic of the deuteron is its anomalous size.
The rms radius,
defined as $r_{\rm rms}=\expected{r^2}^{1/2}$, of the deuteron is 2.8 fm.
That is close to the value for $^{20}$Ne, which has twice the number
of protons and neutrons.

Despite being very weakly bound, $pn$ pairing can undergo the BCS-BEC crossover \cite{Alm1993,Baldo1995,Stein1995}.
For example, this is relevant in heavy-ion collisions where low-density nuclear matter can approximate the state after a collision \cite{Baldo1995}. Another example of $pn$ pairing being relevant is in the case of not completely cooled neutron stars where the fraction of protons is not negligible.

In Ref.~\cite{Baldo1995}, 
the BCS-BEC crossover in nuclear matter was investigated within the BCS framework. The Br\"uckner-Hartree-Fock approach was used to compute single-particle energies $\varepsilon_\textbf{k}$, which then can be used to compute the the gap as a function of the wave-vector, the BCS occupation numbers $n_\textbf{k}$, and the anomalous density (also known as pairing tensor in nuclear physics) in a self-consistent way.
In the low-density limit ($\varepsilon_\textbf{k}$ corresponds to free particles and $n_\textbf{k}$ vanishes) and zero temperature, the gap equation is the Schr\"odinger equation, and the pairing tensor reduces to the deuteron wave function, with the deuteron binding energy as an eigenvalue.

These results can be used to determine the size of the deuteron in matter
\cite{Pistolesi1994}.
If we start at the low-density limit and increase the density, its size decreases until a minimum at $n\sim 0.036$ fm$^{-3}$, after which it grows.

\subsubsection{Asymmetric nuclear matter}
\label{sec:asym}

In Sec.~\ref{sec:sym}, we were dealing with an equal number of protons and neutrons.
However, there are situations where the number of neutrons dramatically exceeds the number of protons, such as proto-neutron stars.
Asymmetric nuclear matter refers to an isospin-imbalanced system, a situation analogous to pairing between a spin-up and spin-down in a polarized (spin-imbalanced) atomic gas.

In Ref.~\cite{Lombardo2001} the
$pn$ pairing in bulk nuclear matter was studied as a function of the density and isospin asymmetry within the BCS framework.
The authors found that in the high-density (weak-coupling) regime, the $pn$ paired state is strongly suppressed even by a minor neutron excess.
As density decreases,
the BCS state with large Cooper pairs evolves smoothly into a BEC of deuterons.
In the resulting low-density state, a neutron excess is not enough to quench the pair correlations because of the large spatial separation of the deuterons and neutrons.
As a result, the deuteron BEC is weakly affected by an additional Fermi sea of the remaining neutrons, even at substantial asymmetries.

\subsubsection{Neutron-neutron pairing}
\label{sec:nn}

In Secs.~\ref{sec:sym} and \ref{sec:asym} we considered systems with two kinds of nucleons, protons and neutrons.
When considering only $nn$ pairing, one is restricted to the BCS side of the crossover since there is no bound state of two neutrons to form a BEC.
The comparison with the BCS side of the crossover of a two-component Fermi gas, Sec.~\ref{sec:twocomp}, is straightforward in this case.

The $nn$ $s$-wave scattering length, $a_{nn}=-18.5$ fm \cite{Gardestig2009}
is substantially larger than the effective range $r^e_{nn}=2.7$ fm.
In the low-density limit, $r^e_{nn} \ll k_F^{-1} \ll |a_{nn}|$, which is the equivalent of the unitary Fermi gas in trapped cold atoms.
However, the finite effective range effects play an important role as we move away from the low-density regime.

The equation of state of a Fermi gas with resonant interactions when the effective range is appreciable
was studied using a $t$-matrix approach in Ref.~\cite{Schwenk2005}, which was then applied to neutron matter.

One of the essential applications of $nn$ pairing is the study of neutron stars, especially the crust.
For a review of the
physics of neutron star crusts, at the
microscopic and macroscopic levels, the reader is referred to Ref.~\cite{Chamel2008}.

It is believed that in the inner crust of a neutron star, there is a dilute gas of neutrons together with a Coulomb lattice of nuclear clusters.
The lattice is composed of ionized atomic nuclei immersed in a nearly constant and uniform charge compensating background of electrons
\cite{Baiko2014}.
As we move toward the center of the star, the neutron gas density goes from 0 to 0.08 fm$^{-1}$. The spacing between the nuclear clusters can be considerable, $\approx$ 100 fm, compared to other relevant length scales.
Hence, the neutron gas can be approximated as uniform matter with corrections due to the Coulomb lattice being of higher order.
The microscopic structure of $nn$ pairing impacts properties that can be observed,
such as the cooling process of neutron stars \cite{Fortin2010}.

Much progress has been made in the low-density regime due to the intrinsic interest and approximations making it more tractable.
The BCS gap equation has been solved in Ref.~\cite{Matsuo2006}, where the authors computed the pairing tensor for several densities.
The wave function resembles a bound state in the low-density limit, but with only a resonance close to zero energy.
The $nn$ pair size was also investigated as a function of the density, which diverges in the vanishing density limit since there is no $nn$ bound state.

Although we discussed many-body systems, up to this point, we centered our discussion around properties of two-body systems: the scattering length and effective range.
Besides the spin/isospin degrees of freedom and the fact that nucleons are fermions, dealing with the nuclear force is complicated because it contains an appreciable three-body component that is not well-understood
(actually, thinking about the nuclear force only in terms of nucleon degrees of freedom is not helpful since mesons play an essential role, the lightest of them being the pions \cite{Madeira2018}).
In Ref.~\cite{Maurizio2014} the authors employed forces obtained from chiral perturbation theory, which include two-body terms and repulsive three-body forces. Although repulsion was introduced in the system, they did not find
a very strong reduction of the gap in neutron matter.
A later study using chiral effective theory also observed the same behavior \cite{Drischler2017}.

Quantum Monte Carlo methods were also used to study neutron matter, not only restricted to the low-density regime. For a detailed review of QMC methods applied to nuclear physics, the reader is referred to Refs.~\cite{Carlson2015} and \cite{Lynn2019}.

One quantity that can be computed straightforwardly in QMC simulations is the pairing gap.
In Ref.~\cite{Gezerlis2010} the authors used Green's function Monte Carlo
(GFMC) to compute the gap using two potential models:
the AV4', a simplified form of the AV18 potential \cite{Wiringa2002}, which includes $s$- and $p$-wave contributions, and the $s$-wave component of AV18 (which makes the potential spherically symmetric and ignores spin-flip terms).
They found that in the low-density regime, the gap's behavior is almost entirely determined by the $s$-wave interaction.
The ratio of the gap to the Fermi energy, $\Delta/E_F$, was consistently lower than the BCS prediction. Its maximum, of $\Delta/E_F \approx$ 0.3, is reached at $-1/(k_F a_{nn})=0.2$, which is very close to the unitary limit ($1/(k_F |a|)=0$).
At higher densities, although the value $k_F a_{nn}$ increases, the finite effective range of the interaction strongly suppresses the gap.
These results agree with the ones obtained with other methods and low-momentum potentials, for example, determinantal QMC simulations using pionless EFT \cite{Abe2009}.

The suppression of the gap concerning the BCS prediction may be a consequence of particle-hole fluctuations.
These screening effects are not present in the BCS formalism but are included in the QMC calculations.
The auxiliary-field diffusion Monte Carlo results of Ref.~\cite{Gandolfi2008},
which employed a more sophisticated potential (AV8' plus a three-body force \cite{Carlson2015}), found almost no suppression of the gap with respect to the BCS result. This is probably due to their choice of the trial wave function.
Although we only discussed a few references, many works that compute neutron matter properties using QMC methods are available \cite{Gezerlis2013,Roggero2014,Wlazlowski2014,Tews2016,Piarulli2020}.

Accounting for screening in theoretical studies is a difficult task. While the nucleon systems are strongly-interacting, polarization is usually included in a perturbative manner.
Despite these complications, progress has been made
\cite{Schwenk2003,Schwenk2004,Cao2006,Zhang2016}.

The possibility to connect the results of both two-component atomic Fermi gases and low-density neutron matter comes from the
low-energy behavior of the phase shift $\delta(k)$, which can be related to $a$ and the effective range $r_e$ \cite{Bethe1949},
\begin{eqnarray}
\label{eq:phase_shifts}
k \cot \delta(k)=-\frac{1}{a}+\frac{r_e k^2}{2}+\mathcal{O}(k^4).
\end{eqnarray}
This equation is often called shape independent approximation because
different potentials that reproduce the same scattering length and effective range yield the same low-energy phase-shift behavior.
In dilute cold gases, the effective range $r_e$ between atoms is
much smaller than the interatomic spacing $r_0$, and can be taken to be
zero. The diluteness can guarantee
that the scattering length $a$ is much larger than $r_0$. Comparison
with other systems is meaningful if they also obey 
$|a| \gg r_0 \gg r_e$.
The scattering length of neutron matter is
substantially larger than the interparticle distance and the effective range,
such that $|r_e^{nn}/a^{nn}|\approx 0.15$.
However, only at very low-densities is the effective range much smaller
than the interparticle distance.  If we neglect the effects of a finite
effective range in the neutron-neutron interaction, cold atoms, and neutron
matter are universal in the sense that properties depend only on the product
$k_F a$.

\begin{figure}[!htb]
\centering
  \includegraphics[width=\linewidth]{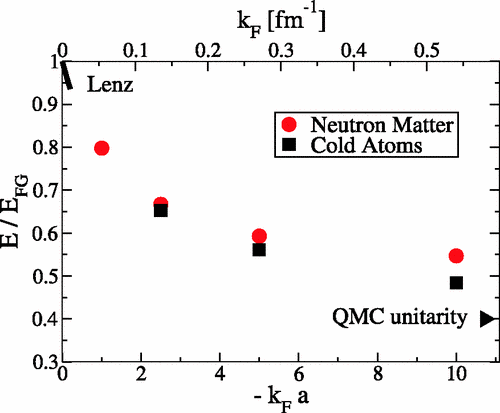}
\caption{
Equation of state for cold
atoms and neutron matter at zero temperature computed using QMC methods.
The triangle shows the cold atoms result at unitarity.
The bottom $x$-axis corresponds to the interaction strength $-k_F a$, while the top $x$-axis shows the value of $k_F$ if we substitute the value for the $nn$ scattering length, $a_{nn}=-18.5$ fm.
Notice that, as the density decreases, the energy of both systems becomes closer together.
Taken from Ref.~\cite{Gezerlis2008}}
\label{fig:2}       % Give a unique label
\end{figure}

In Ref.~\cite{Gezerlis2008}, the authors used QMC methods to compute the equation of state of both cold atoms and low-density neutron matter, and they found both to be very similar, Fig.~\ref{fig:2}.
They also computed the neutron matter pairing gap, which is significantly suppressed relative to cold atoms. The difference was attributed to the finite effective
range in the neutron-neutron interaction.
Reference \cite{Madeira2019}, besides confirming these results, considered vortex-line excitations to these fluids.
The authors also found agreement in some vortex properties
between cold gases and neutron matter for very low densities.
However,
the density depletion at the vortex core, which depends strongly on the short-ranged interaction cold atomic gases, is approximately constant for neutron matter in the low-density regime.

\subsubsection{Alpha particle condensation}
\label{sec:alpha}

In Sec.~\ref{sec:sym}, we covered the BCS-BEC crossover in systems with an equal number of protons and neutrons, where the Cooper pairs were $pn$ and the BEC state corresponded to the deuteron.
This is the description that comes to mind when we draw the parallel with two-component Fermi gases.
However, nucleons correspond to four states due to the spin/isospin degrees of freedom.
Hence we can consider the possibility of quartet condensation, where the four-fermion object is the $\alpha$ particle
(a bound state of two protons and two neutrons, the $^4$He nucleus).

This topic has not been so extensively covered as the conventional BCS-BEC crossover because of the technical difficulties that arise, both in theoretical and numerical approaches.
However, a quartet phase has been predicted in theoretical studies
\cite{Capponi2007,Capponi2008}.
Also, QMC simulations of
four-component fermions with unitary interactions found
a ground state of the eight-particle system whose energy is almost equal to that of two four-particle systems \cite{Dawkins2020}.

Also, the BCS-BEC crossover for quartets is not qualitatively the same as the usual one: for high densities the quartets break up into two Cooper pairs.
The theoretical description of quartet condensation in Fermi systems, with applications to nuclei and nuclear matter, has been discussed in Refs.~\cite{Kamei2005,Sogo2009,Sogo2010,Sogo2010b,Schuck2014}.

\subsubsection{Nuclei}

Nuclei are self-bound, which means that the attractive interactions inside a nucleus overcome the repulsive ones to form a cluster. In cold atoms, an external potential can be applied to confine systems even with purely repulsive interactions \cite{Bloch2008,Giorgini2008}.
Although in Secs.~\ref{sec:sym} to \ref{sec:alpha} we talked about nuclei, they were immersed in infinite matter.
We can also discuss the BCS-BEC crossover and alpha condensation in nuclei in vacuum.
For heavy nuclei, where the number of neutrons exceeds the number of protons, $pn$ pairing is suppressed because of the same reasons we listed in Sec.~\ref{sec:asym}. However, in nuclei where the number of neutrons and protons is comparable, $pn$ could be possible because this channel's attraction exceeds the $pp$ or $nn$ ones. References~\cite{Gezerlis2011} and \cite{Bulthuis2016} address this subject.

We can also consider viewing some light nuclei as a cluster plus a single Cooper pair to investigate $pn$ pairing.
A few words about the terminology are in order. It may seem strange to refer to a single Cooper pair. Still, we intend to discuss pairing between nucleons as a function of the interaction strength and draw a parallel with what we already saw in infinite nuclear matter and cold atoms.
In this sense, the BEC state is simply the deuteron.
Going back to our discussion, $^6$Li, $^{18}$F, and $^{42}$Sc are examples of nuclei with a structure ``cluster plus deuteron'' (the clusters being $\alpha$, $^{16}$O, $^{40}$Ca, respectively).
Because of the relatively small number of nucleons, results under this perspective are only available for $^6$Li \cite{Kamimura1981}.
Much like the infinite matter case, the size of the deuteron (as a function of the distance to the $\alpha$) diminishes before it enters the $\alpha$ particle.

Neutron-neutron pairing is also important to understand the properties of neutron-rich nuclei.
In heavy nuclei, close to the neutron drip line, a neutron-rich outer layer called neutron skin is present.
Interestingly, the $nn$ Cooper pairs are localized on the surface of these nuclei.
Details can be found in Refs.~\cite{Matsuo2005,Pillet2007,Pillet2010,Hagino2010}.

As for the BEC of $\alpha$ particles in nuclei, usually the ground-state is too small to allow $\alpha$ condensation (an exception is $^8$Be, which shows a two $\alpha$ cluster structure \cite{Wiringa2000}).
However, there are long-lived excited states of light nuclei that show clustering of $\alpha$ particles. One of these states is the famous Hoyle state \cite{Hoyle1954},
an excited state of $^{12}$C produced by a triple-$\alpha$ process,
essential for the nucleosynthesis of carbon in stars.
For a more in-depth explanation, and some other candidates for $\alpha$ condensation, the reader is referred to Refs.~\cite{Yamada2012,Schuck2016,Tohsaki2017}.

\section{Free expansion of a turbulent quantum cloud of cold atoms: the merging of turbulence with statistics of matter-waves}
\label{sec:laser}

Statistical optics is a field of physics that has attracted much interest recently.
Several forms of analysis regarding photons and their distributions are part of this area's interests, among these, the so-called speckle fields.
Light waves belonging to a coherent beam, such as a laser beam, scattered over surfaces containing randomly distributed scatters, generate interference patterns in the light fields called speckles.
Speckles constitute a random collection of amplitudes and phases that reveal a wide variety of light beams' properties. They have applications in several fields of science and technology.

Although the speckles' field is largely focused on light waves, we can equivalently represent them as matter-waves.
This would undoubtedly represent the inclusion of matter-waves in an optical statistics framework. In fact, with the advent of atomic BECs, we can think about these possibilities more easily than before.
When we produce a condensate and let it expand freely in space, what we have is a coherent beam of matter, precisely equivalent to a laser beam for propagating electromagnetic waves. If we introduce a collection of random amplitudes and phases into this condensate, we will have the equivalent of a field of speckle of matter-waves there. 

To establish this parallel, Hussein and collaborators \cite{Tavares2017} envisioned a turbulent condensate production and expansion.
An unperturbed condensate is a superfluid.
With the introduction of excitations, which can generate vortices and waves in this superfluid, quantum turbulence can be reached \cite{Madeira2020,Madeira2020b}.
Once in this state, we have many amplitude and phase fluctuations, reaching a situation very similar to that expected in speckle fields.

The turbulent regime was achieved experimentally and characterized. Once in this state, the system was released from the trap, corresponding to a matter-wave propagating in space. Still, instead of having the characteristics of a coherent and well-behaved wave, it was a collection of coherent domains, much like in speckle fields.

The researchers analyzed mainly two characteristics of the spatial disorder of the systems. First, measurements of the aspect ratios of regular and turbulent BECs were performed.
It is known that for ground-state BECs there is an inversion of the aspect ratio of the cloud in time-of-flight (TOF) measurements, whereas in the turbulent
case there is a self-similar expansion, without ever inverting its
aspect ratio \cite{Caracanhas2013}.
For a coherent Gaussian beam,
there is also an inversion of the aspect ratio of the waists, whereas it is
preserved in the propagation of the elliptical speckle light map.

\begin{figure*}[!htb]
\centering
  \includegraphics[width=0.45\linewidth]{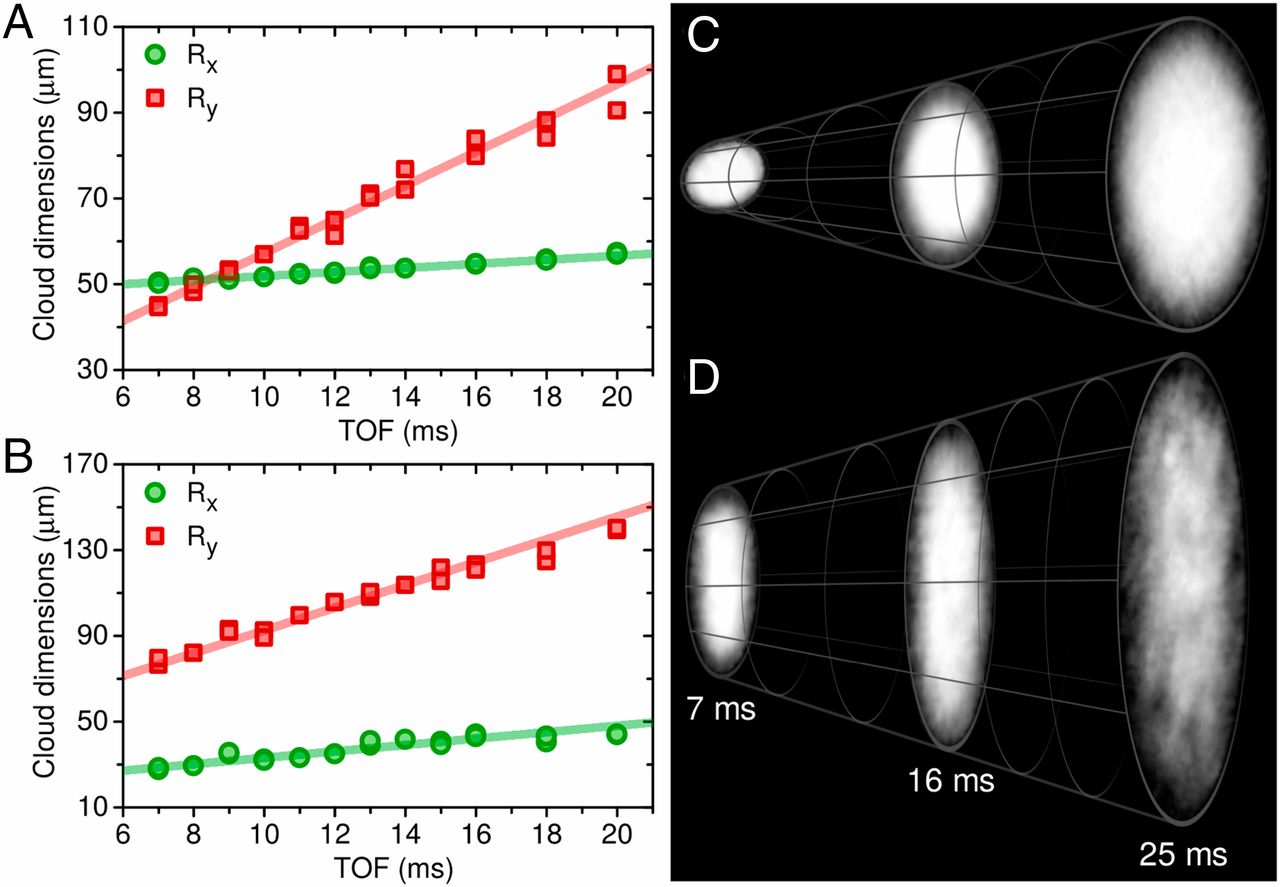}
  \includegraphics[width=0.45\linewidth]{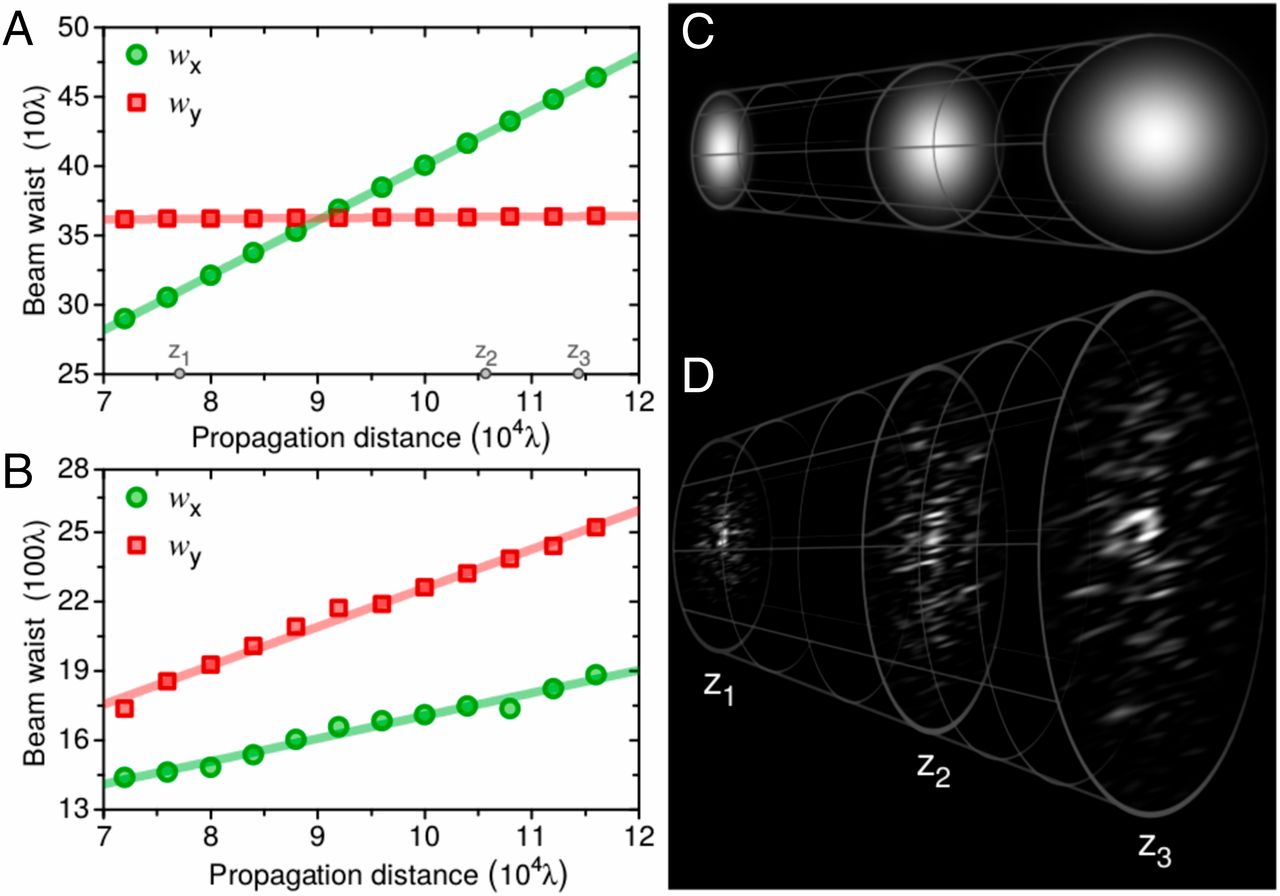}
\caption{
Comparison between BECs and optical beams.
In the left panel, the BEC expansion during TOF is investigated. The Thomas-Fermi radii as a function of the TOF are plotted for standard (A) and turbulent (B) BECs. Then, three different TOF snapshots showing the expansion of standard (C) and turbulent BECs (D) are shown.
The panel on the right concerns the
optical beam propagation. The optical beam waists as a function of the axial distance are plotted for a coherent Gaussian beam (A) and the speckle beam (B). A sequence of three different propagation distances shows the expansion of a Gaussian beam (C) and a speckle beam (D).
At first glance, the similarity between the plots is intriguing, but they can be understood under the light of both being coherent matter-wave systems.
Taken from Ref.~\cite{Tavares2017}.}
\label{fig:3}       % Give a unique label
\end{figure*}

The analogy between the two systems is constructed through
the state of disorder that characterizes the matter-waves.
In Fig.~\ref{fig:3}, we show the behavior of the propagation of an unperturbed condensate and its characteristics.
When expanding, the behavior is like that found in a beam of Gaussian laser light: as it propagates, the smaller dimensions undergo a more significant divergence due to diffraction, resulting in an inversion of the so-called aspect-ratio as the propagation progresses.
This is exactly what is found in the propagation of a beam of light containing a speckle field. Due to the domains, diffraction is no longer dominated by the wavelength to the beam ratio, but by its correlation length.

The second property that was investigated was the coherence in both
systems.
It was found that the correlations in regular BECs resemble the ones in
the Gaussian beam, while the same is true for the turbulent BEC and speckle
beam pair.

This was the first experimental evidence of a three-dimensional speckle matter-wave. The production of three-dimensional speckle fields is an exciting problem to be considered, and here, with the turbulent conditions and expansion, we naturally have this connection. Even more interestingly is the fact that the study of speckle fields as matter-waves and their statistical behavior can now be considered. This certainly opens the opportunity of promoting the merging of quantum turbulence with statistical optics, now in the context of matter-waves.

\section{Conclusions and outlook}
\label{sec:conc}

In conclusion, we pointed out similarities and differences between cold atom systems and nuclear physics, and also statistical optics.
These topics where chosen because of the interesting comparisons that can be made between cold atoms and them. A full list of all the systems that similarities could be investigated would include superconductors, quantum chromodynamics (QCD), and many other topics beyond this review's scope.

Concerning nuclear physics, we chose to center the discussion around the BCS-BEC crossover because we believe it is a very clear and appropriate comparison. However, we did not cover all the related topics, for example, finite temperature effects \cite{Strinati2018}.

Besides the crossover, we could have centered the discussion on other characteristics that connect cold atoms and nuclear physics, such as:
the role of three-body forces \cite{Hammer2013} (specially under the perspective of effective field theories \cite{Kolck1994,Kolck1999,Friar1999,Bedaque2002}),
the finite-size character \cite{Zinner2013},
the Efimov effect \cite{Braaten2007,Naidon2017}, and halo nuclei \cite{Frederico2012a}, to cite a few examples.

Also, much of our discussion concerned two-component Fermi gases because of the clear comparison with neutron matter.
Three-component fermionic gases have been produced \cite{Ottenstein2008,Huckans2009,Nakajima2010}, which increases our understanding of multi-component Fermi systems.
However, we believe a significant breakthrough will occur when four-component fermionic gases can be produced experimentally and properties calculated from theoretical models. Nuclear physics is a four-state fermionic problem (protons and neutrons with two possibilities for the spin), and being able to produce analogous systems in the cold atoms scenario would be illuminating.

The second parallel we drew was between BECs and statistical optics, and it was used to study quantum turbulence.
Turbulence is characterized by a large number of degrees of freedom, distributed over several length scales, that result in a disordered state of a fluid. The field of quantum turbulence deals with the manifestation of turbulence in quantum fluids, such as trapped cold gases.
There are some inherent challenges in determining and characterizing the turbulent state in these systems \cite{Madeira2020,Madeira2020b}.
Hence, different approaches have been used to clarify aspects of quantum turbulence \cite{Orozco2020,Marino2020,Madeira2020c}.
What we showed in Sec.~\ref{sec:laser} can be used to look at the phenomena from a different perspective, which is of paramount importance when dealing with quantum turbulence.

It is not uncommon to find works where the connection between cold atoms and another field is mentioned in the introduction or outlook.
However, there are not many that make a serious effort to compare and contrast them. If that were to change, we believe that this would be beneficial to both areas.

\begin{acknowledgements}
At first glance, the parallel between BECs and statistical optics may seem slightly out of place.
However, one of the authors of Ref.~\cite{Tavares2017} is
Mahir Saleh Hussein.
Two of his greatest interests were nuclear physics and Bose-Einstein condensation \cite{Balantekin2019}.
Since this is a special issue in his honor, we thought this two-fold tribute would be appropriate.
This work was supported by
the S\~ao Paulo Research Foundation (FAPESP)
under the grants 2013/07276-1, 2014/50857-8, and 2018/09191-7, and by the
National Council for Scientific and Technological Development (CNPq)
under the grant 465360/2014-9.
\end{acknowledgements}

\section*{Conflict of interest}

The authors declare that they have no conflict of interest.

\bibliographystyle{spphys}       % APS-like style for physics
\bibliography{madeira_v2}

\begin{thebibliography}{100}
\providecommand{\url}[1]{{#1}}
\providecommand{\urlprefix}{URL }
\expandafter\ifx\csname urlstyle\endcsname\relax
  \providecommand{\doi}[1]{DOI \discretionary{}{}{}#1}\else
  \providecommand{\doi}{DOI \discretionary{}{}{}\begingroup
  \urlstyle{rm}\Url}\fi

\bibitem{Anderson1995}
M.H. Anderson, J.R. Ensher, M.R. Matthews, C.E. Wieman, E.A. Cornell, Science
  \textbf{269}(5221), 198 (1995)

\bibitem{Bradley1995}
C.C. Bradley, C.A. Sackett, J.J. Tollett, R.G. Hulet, Phys. Rev. Lett.
  \textbf{75}(9), 1687 (1995)

\bibitem{Davis1995}
K.B. Davis, M.O. Mewes, M.R. Andrews, N.J. {Van Druten}, D.S. Durfee, D.M.
  Kurn, W.~Ketterle, Phys. Rev. Lett. \textbf{75}(22), 3969 (1995)

\bibitem{Ohara2002}
K.M. O'Hara, S.L. Hemmer, M.E. Gehm, S.R. Granade, J.E. Thomas, Science
  \textbf{298}(5601), 2179 (2002)

\bibitem{Zwierlein2005}
M.W. Zwierlein, J.R. Abo-Shaeer, A.~Schirotzek, C.H. Schunck, W.~Ketterle,
  Nature \textbf{435}(7045), 1047 (2005)

\bibitem{Pethick2008}
C.J. Pethick, H.~Smith, \emph{{Bose-Einstein condensation in dilute gases}},
  vol. 9780521846 (Cambridge University Press, 2008)

\bibitem{Zinner2013}
N.T. Zinner, A.S. Jensen, J. Phys. G Nucl. Part. Phys. \textbf{40}(5), 053101
  (2013)

\bibitem{Amorim1997}
A.E. Amorim, T.~Frederico, L.~Tomio, Phys. Rev. C - Nucl. Phys. \textbf{56}(5),
  R2378 (1997)

\bibitem{Riisager2000}
K.~Riisager, D.V. Fedorov, A.S. Jensen, Europhys. Lett. \textbf{49}(5), 547
  (2000)

\bibitem{Bardeen1957}
J.~Bardeen, L.N. Cooper, J.R. Schrieffer, Phys. Rev. \textbf{108}(5), 1175
  (1957)

\bibitem{Bose1924}
Bose, Zeitschrift f{\"{u}}r Phys. \textbf{26}(1), 178 (1924)

\bibitem{Strinati2018}
G.C. Strinati, P.~Pieri, G.~R{\"{o}}pke, P.~Schuck, M.~Urban, Phys. Rep.
  \textbf{738}, 1 (2018)

\bibitem{Tavares2017}
P.E.S. Tavares, A.R. Fritsch, G.D. Telles, M.S. Hussein, F.~Impens, R.~Kaiser,
  V.S. Bagnato, Proc. Natl. Acad. Sci. \textbf{114}(48), 12691 (2017)

\bibitem{Madeira2020}
L.~Madeira, M.~Caracanhas, F.~dos Santos, V.~Bagnato, Annu. Rev. Condens.
  Matter Phys. \textbf{11}(1), 37 (2020)

\bibitem{Madeira2020b}
L.~Madeira, A.~Cidrim, M.~Hemmerling, M.A. Caracanhas, F.E.A. dos Santos, V.S.
  Bagnato, AVS Quantum Sci. \textbf{2}(3), 035901 (2020)

\bibitem{Bloch2008}
I.~Bloch, J.~Dalibard, W.~Zwerger, Rev. Mod. Phys. \textbf{80}(3), 885 (2008)

\bibitem{Giorgini2008}
S.~Giorgini, L.P. Pitaevskii, S.~Stringari, Rev. Mod. Phys. \textbf{80}(4),
  1215 (2008)

\bibitem{Onofrio2016}
R.~Onofrio, Physics-Uspekhi \textbf{59}(11), 1129 (2016)

\bibitem{Ketterle2008}
W.~Ketterle, M.W. Zwierlein, Riv. del Nuovo Cim.  (2008)

\bibitem{Ottenstein2008}
T.B. Ottenstein, T.~Lompe, M.~Kohnen, A.N. Wenz, S.~Jochim, Phys. Rev. Lett.
  \textbf{101}(20), 203202 (2008)

\bibitem{Huckans2009}
J.H. Huckans, J.R. Williams, E.L. Hazlett, R.W. Stites, K.M. O'Hara, Phys. Rev.
  Lett. \textbf{102}(16), 165302 (2009)

\bibitem{Nakajima2010}
S.~Nakajima, M.~Horikoshi, T.~Mukaiyama, P.~Naidon, M.~Ueda, Phys. Rev. Lett.
  \textbf{105}(2), 023201 (2010)

\bibitem{Zwerger2011}
W.~Zwerger, \emph{{The BCS-BEC Crossover and the Unitary Fermi Gas}},
  \emph{Lecture Notes in Physics}, vol. 836 (Springer Berlin Heidelberg,
  Berlin, Heidelberg, 2012)

\bibitem{Randeria2014}
M.~Randeria, E.~Taylor, Annu. Rev. Condens. Matter Phys. \textbf{5}(1), 209
  (2014)

\bibitem{Eagles1969}
D.M. Eagles, Phys. Rev. \textbf{186}(2), 456 (1969)

\bibitem{Leggett2008}
A.J. Leggett, in \emph{Mod. Trends Theory Condens. Matter} (Springer Berlin
  Heidelberg, 2008), pp. 13--27

\bibitem{Regal2004}
C.A. Regal, M.~Greiner, D.S. Jin, Phys. Rev. Lett. \textbf{92}(4), 4 (2004)

\bibitem{Moerdijk1995}
A.J. Moerdijk, B.J. Verhaar, A.~Axelsson, Phys. Rev. A \textbf{51}(6), 4852
  (1995)

\bibitem{Baker1999}
G.A. Baker, Phys. Rev. C \textbf{60}(5), 054311 (1999)

\bibitem{Carlson2011}
J.~Carlson, S.~Gandolfi, K.E. Schmidt, S.~Zhang, Phys. Rev. A - At. Mol. Opt.
  Phys. \textbf{84}(6), 061602 (2011)

\bibitem{Ku2012}
M.J. Ku, A.T. Sommer, L.W. Cheuk, M.W. Zwierlein, Science \textbf{335}(6068),
  563 (2012)

\bibitem{Zurn2013}
G.~Z{\"{u}}rn, T.~Lompe, A.N. Wenz, S.~Jochim, P.S. Julienne, J.M. Hutson,
  Phys. Rev. Lett. \textbf{110}(13), 135301 (2013)

\bibitem{Foulkes2001}
W.M. Foulkes, L.~Mitas, R.J. Needs, G.~Rajagopal, Rev. Mod. Phys.
  \textbf{73}(1), 33 (2001)

\bibitem{Carlson2012}
J.~Carlson, S.~Gandolfi, A.~Gezerlis, Prog. Theor. Exp. Phys. \textbf{2012}(1)
  (2012)

\bibitem{Carlson2015}
J.~Carlson, S.~Gandolfi, F.~Pederiva, S.C. Pieper, R.~Schiavilla, K.E. Schmidt,
  R.B. Wiringa, Rev. Mod. Phys. \textbf{87}(3), 1067 (2015)

\bibitem{Lynn2019}
J.~Lynn, I.~Tews, S.~Gandolfi, A.~Lovato, Annu. Rev. Nucl. Part. Sci.
  \textbf{69}(1), 279 (2019)

\bibitem{Carlson2003}
J.~Carlson, S.Y. Chang, V.R. Pandharipande, K.E. Schmidt, Phys. Rev. Lett.
  \textbf{91}(5), 050401 (2003)

\bibitem{Astrakharchik2004}
G.E. Astrakharchik, J.~Boronat, J.~Casulleras, S.~Giorgini, Phys. Rev. Lett.
  \textbf{93}(20), 200404 (2004)

\bibitem{Chang2004}
S.Y. Chang, V.R. Pandharipande, J.~Carlson, K.E. Schmidt.
\newblock {Quantum Monte Carlo studies of superfluid Fermi gases} (2004)

\bibitem{Gandolfi2011b}
S.~Gandolfi, K.E. Schmidt, J.~Carlson, Phys. Rev. A \textbf{83}(4), 041601
  (2011)

\bibitem{Bulgac2012}
A.~Bulgac, M.M. Forbes, P.~Magierski, in \emph{Lect. Notes Phys.}, vol. 836
  (Springer, Berlin, Heidelberg, 2012), pp. 305--373

\bibitem{Carlson2013}
J.~Carlson, S.~Gandolfi, A.~Gezerlis, in \emph{Fifty Years Nucl. BCS} (World
  Scientific, 2013), pp. 348--359

\bibitem{Gandolfi2014}
S.~Gandolfi, J. Phys. Conf. Ser. \textbf{529}, 012011 (2014)

\bibitem{Forbes2011}
M.M. Forbes, S.~Gandolfi, A.~Gezerlis, Phys. Rev. Lett. \textbf{106}(23),
  235303 (2011)

\bibitem{Forbes2012}
M.M. Forbes, S.~Gandolfi, A.~Gezerlis, Phys. Rev. A \textbf{86}(5), 053603
  (2012)

\bibitem{Pessoa2015}
R.~Pessoa, S.~Gandolfi, S.A. Vitiello, K.E. Schmidt, Phys. Rev. A
  \textbf{92}(6), 063625 (2015)

\bibitem{Pessoa2015b}
R.~Pessoa, S.A. Vitiello, K.E. Schmidt, J. Low Temp. Phys. \textbf{180}(1-2),
  168 (2015)

\bibitem{Pessoa2019}
R.~Pessoa, S.A. Vitiello, K.E. Schmidt, Phys. Rev. A \textbf{100}(5), 053601
  (2019)

\bibitem{Hoinka2013}
S.~Hoinka, M.~Lingham, K.~Fenech, H.~Hu, C.J. Vale, J.E. Drut, S.~Gandolfi,
  Phys. Rev. Lett. \textbf{110}(5), 055305 (2013)

\bibitem{Galea2016}
A.~Galea, H.~Dawkins, S.~Gandolfi, A.~Gezerlis, Phys. Rev. A \textbf{93}(2),
  023602 (2016)

\bibitem{Galea2017}
A.~Galea, T.~Zielinski, S.~Gandolfi, A.~Gezerlis, J. Low Temp. Phys.
  \textbf{189}(5-6), 451 (2017)

\bibitem{Madeira2016}
L.~Madeira, S.A. Vitiello, S.~Gandolfi, K.E. Schmidt, Phys. Rev. A
  \textbf{93}(4), 043604 (2016)

\bibitem{Madeira2019}
L.~Madeira, S.~Gandolfi, K.E. Schmidt, V.S. Bagnato, Phys. Rev. C
  \textbf{100}(1), 014001 (2019)

\bibitem{Madeira2017}
L.~Madeira, S.~Gandolfi, K.E. Schmidt, Phys. Rev. A \textbf{95}(5), 053603
  (2017)

\bibitem{Gezerlis2009}
A.~Gezerlis, S.~Gandolfi, K.E. Schmidt, J.~Carlson, Phys. Rev. Lett.
  \textbf{103}(6), 060403 (2009)

\bibitem{Alm1993}
T.~Alm, B.L. Friman, G.~R{\"{o}}pke, H.~Schulz, Nucl. Physics, Sect. A
  \textbf{551}(1), 45 (1993)

\bibitem{Baldo1995}
M.~Baldo, U.~Lombardo, P.~Schuck, Phys. Rev. C \textbf{52}(2), 975 (1995)

\bibitem{Stein1995}
H.~Stein, A.~Schnell, T.~Alm, G.~R{\"{o}}pk, Zeitschrift f{\"{u}}r Phys. A
  Hadron. Nucl. \textbf{351}(3), 295 (1995)

\bibitem{Pistolesi1994}
F.~Pistolesi, G.C. Strinati, Phys. Rev. B \textbf{49}(9), 6356 (1994)

\bibitem{Lombardo2001}
U.~Lombardo, P.~Nozi{\`{e}}res, P.~Schuck, H.J. Schulze, A.~Sedrakian, Phys.
  Rev. C - Nucl. Phys. \textbf{64}(6), 643141 (2001)

\bibitem{Gardestig2009}
A.~G{\aa}rdestig, J. Phys. G Nucl. Part. Phys. \textbf{36}(5), 053001 (2009)

\bibitem{Schwenk2005}
A.~Schwenk, C.J. Pethick, Phys. Rev. Lett. \textbf{95}(16), 160401 (2005)

\bibitem{Chamel2008}
N.~Chamel, P.~Haensel, Living Rev. Relativ. \textbf{11}(1), 10 (2008)

\bibitem{Baiko2014}
D.A. Baiko, J. Phys. Conf. Ser. \textbf{496}(1), 012010 (2014)

\bibitem{Fortin2010}
M.~Fortin, F.~Grill, J.~Margueron, D.~Page, N.~Sandulescu, Phys. Rev. C - Nucl.
  Phys. \textbf{82}(6), 065804 (2010)

\bibitem{Matsuo2006}
M.~Matsuo, Phys. Rev. C - Nucl. Phys. \textbf{73}(4), 044309 (2006)

\bibitem{Madeira2018}
L.~Madeira, A.~Lovato, F.~Pederiva, K.E. Schmidt, Phys. Rev. C \textbf{98}(3),
  034005 (2018)

\bibitem{Maurizio2014}
S.~Maurizio, J.W. Holt, P.~Finelli, Phys. Rev. C - Nucl. Phys. \textbf{90}(4),
  044003 (2014)

\bibitem{Drischler2017}
C.~Drischler, T.~Kr{\"{u}}ger, K.~Hebeler, A.~Schwenk, Phys. Rev. C
  \textbf{95}(2), 024302 (2017)

\bibitem{Gezerlis2010}
A.~Gezerlis, J.~Carlson, Phys. Rev. C \textbf{81}(2), 025803 (2010)

\bibitem{Wiringa2002}
R.B. Wiringa, S.C. Pieper, Phys. Rev. Lett. \textbf{89}(18), 182501 (2002)

\bibitem{Abe2009}
T.~Abe, R.~Seki, Phys. Rev. C - Nucl. Phys. \textbf{79}(5), 054002 (2009)

\bibitem{Gandolfi2008}
S.~Gandolfi, A.Y. Illarionov, S.~Fantoni, F.~Pederiva, K.E. Schmidt, Phys. Rev.
  Lett. \textbf{101}(13), 132501 (2008)

\bibitem{Gezerlis2013}
A.~Gezerlis, I.~Tews, E.~Epelbaum, S.~Gandolfi, K.~Hebeler, A.~Nogga,
  A.~Schwenk, Phys. Rev. Lett. \textbf{111}(3), 032501 (2013)

\bibitem{Roggero2014}
A.~Roggero, A.~Mukherjee, F.~Pederiva, Phys. Rev. Lett. \textbf{112}(22),
  221103 (2014)

\bibitem{Wlazlowski2014}
G.~Wlaz{\l}owski, J.W. Holt, S.~Moroz, A.~Bulgac, K.J. Roche, Phys. Rev. Lett.
  \textbf{113}(18), 182503 (2014)

\bibitem{Tews2016}
I.~Tews, S.~Gandolfi, A.~Gezerlis, A.~Schwenk, Phys. Rev. C \textbf{93}(2),
  024305 (2016)

\bibitem{Piarulli2020}
M.~Piarulli, I.~Bombaci, D.~Logoteta, A.~Lovato, R.B. Wiringa, Phys. Rev. C
  \textbf{101}(4), 045801 (2020)

\bibitem{Schwenk2003}
A.~Schwenk, B.~Friman, G.E. Brown, Nucl. Phys. A \textbf{713}(1-2), 191 (2003)

\bibitem{Schwenk2004}
A.~Schwenk, B.~Friman, Phys. Rev. Lett. \textbf{92}(8), 082501 (2004)

\bibitem{Cao2006}
L.G. Cao, U.~Lombardo, P.~Schuck, Phys. Rev. C - Nucl. Phys. \textbf{74}(6),
  064301 (2006)

\bibitem{Zhang2016}
S.S. Zhang, L.G. Cao, U.~Lombardo, P.~Schuck, Phys. Rev. C \textbf{93}(4),
  044329 (2016)

\bibitem{Bethe1949}
H.A. Bethe, Phys. Rev. \textbf{76}(1), 38 (1949)

\bibitem{Gezerlis2008}
A.~Gezerlis, J.~Carlson, Phys. Rev. C \textbf{77}(3), 032801 (2008)

\bibitem{Capponi2007}
S.~Capponi, G.~Roux, P.~Azaria, E.~Boulat, P.~Lecheminant, Phys. Rev. B -
  Condens. Matter Mater. Phys. \textbf{75}(10), 100503 (2007)

\bibitem{Capponi2008}
S.~Capponi, G.~Roux, P.~Lecheminant, P.~Azaria, E.~Boulat, S.R. White, Phys.
  Rev. A - At. Mol. Opt. Phys. \textbf{77}(1), 013624 (2008)

\bibitem{Dawkins2020}
W.G. Dawkins, J.~Carlson, U.~{Van Kolck}, A.~Gezerlis, Phys. Rev. Lett.
  \textbf{124}(14), 143402 (2020)

\bibitem{Kamei2005}
H.~Kamei, K.~Miyake, J. Phys. Soc. Japan \textbf{74}(7), 1911 (2005)

\bibitem{Sogo2009}
T.~Sogo, R.~Lazauskas, G.~R{\"{o}}pke, P.~Schuck, Phys. Rev. C - Nucl. Phys.
  \textbf{79}(5), 051301 (2009)

\bibitem{Sogo2010}
T.~Sogo, G.~R{\"{o}}pke, P.~Schuck, Phys. Rev. C - Nucl. Phys. \textbf{81}(6),
  064310 (2010)

\bibitem{Sogo2010b}
T.~Sogo, G.~R{\"{o}}pke, P.~Schuck, Phys. Rev. C - Nucl. Phys. \textbf{82}(3),
  034322 (2010)

\bibitem{Schuck2014}
P.~Schuck, Y.~Funaki, H.~Horiuchi, G.~R{\"{o}}pke, A.~Tohsaki, T.~Yamada, J.
  Phys. Conf. Ser. \textbf{529}(1), 012014 (2014)

\bibitem{Gezerlis2011}
A.~Gezerlis, G.F. Bertsch, Y.L. Luo, Phys. Rev. Lett. \textbf{106}(25), 252502
  (2011)

\bibitem{Bulthuis2016}
B.~Bulthuis, A.~Gezerlis, Phys. Rev. C \textbf{93}(1), 014312 (2016)

\bibitem{Kamimura1981}
M.~Kamimura, Nucl. Physics, Sect. A \textbf{351}(3), 456 (1981)

\bibitem{Matsuo2005}
M.~Matsuo, K.~Mizuyama, Y.~Serizawa, Phys. Rev. C - Nucl. Phys. \textbf{71}(6),
  064326 (2005)

\bibitem{Pillet2007}
N.~Pillet, N.~Sandulescu, P.~Schuck, Phys. Rev. C - Nucl. Phys. \textbf{76}(2),
  024310 (2007)

\bibitem{Pillet2010}
N.~Pillet, N.~Sandulescu, P.~Schuck, J.F. Berger, Phys. Rev. C - Nucl. Phys.
  \textbf{81}(3), 034307 (2010)

\bibitem{Hagino2010}
K.~Hagino, H.~Sagawa, P.~Schuck, J. Phys. G Nucl. Part. Phys. \textbf{37}(6),
  064040 (2010)

\bibitem{Wiringa2000}
R.B. Wiringa, S.C. Pieper, J.~Carlson, V.R. Pandharipande, Phys. Rev. C - Nucl.
  Phys. \textbf{62}(1), 23 (2000)

\bibitem{Hoyle1954}
F.~Hoyle, Astrophys. J. Suppl. Ser. \textbf{1}, 121 (1954)

\bibitem{Yamada2012}
T.~Yamada, Y.~Funaki, H.~Horiuchi, G.~R{\"{o}}pke, P.~Schuck, A.~Tohsaki,
  \emph{{Nuclear Alpha-Particle Condensates}} (Springer Berlin Heidelberg,
  Berlin, Heidelberg, 2012), pp. 229--298

\bibitem{Schuck2016}
P.~Schuck, Y.~Funaki, H.~Horiuchi, G.~R{\"{o}}pke, A.~Tohsaki, T.~Yamada, Phys.
  Scr. \textbf{91}(12), 123001 (2016)

\bibitem{Tohsaki2017}
A.~Tohsaki, H.~Horiuchi, P.~Schuck, G.~R{\"{o}}pke, Rev. Mod. Phys.
  \textbf{89}(1), 011002 (2017)

\bibitem{Caracanhas2013}
M.~Caracanhas, A.L. Fetter, G.~Baym, S.R. Muniz, V.S. Bagnato, J. Low Temp.
  Phys. \textbf{170}(3-4), 133 (2013)

\bibitem{Hammer2013}
H.W. Hammer, A.~Nogga, A.~Schwenk, Rev. Mod. Phys. \textbf{85}(1), 197 (2013)

\bibitem{Kolck1994}
U.~{Van Kolck}, Phys. Rev. C \textbf{49}(6), 2932 (1994)

\bibitem{Kolck1999}
U.~{Van Kolck}, Nucl. Phys. A \textbf{645}(2), 273 (1999)

\bibitem{Friar1999}
J.L. Friar, D.~H{\"{u}}ber, U.~van Kolck, Phys. Rev. C - Nucl. Phys.
  \textbf{59}(1), 53 (1999)

\bibitem{Bedaque2002}
P.F. Bedaque, U.~van Kolck, Annu. Rev. Nucl. Part. Sci. \textbf{52}(1), 339
  (2002)

\bibitem{Braaten2007}
E.~Braaten, H.W. Hammer, Ann. Phys. (N. Y). \textbf{322}(1), 120 (2007)

\bibitem{Naidon2017}
P.~Naidon, S.~Endo, Reports Prog. Phys. \textbf{80}(5), 056001 (2017)

\bibitem{Frederico2012a}
T.~Frederico, A.~Delfino, L.~Tomio, M.T. Yamashita, Prog. Part. Nucl. Phys.
  \textbf{67}(4), 939 (2012)

\bibitem{Orozco2020}
A.~{Daniel Garc{\'{i}}a-Orozco}, L.~Madeira, L.~Galantucci, C.F. Barenghi, V.S.
  Bagnato, EPL (Europhysics Lett. \textbf{130}(4), 46001 (2020)

\bibitem{Marino2020}
{\'{A}}.V.M. Marino, L.~Madeira, A.~Cidrim, F.E.A. dos Santos, V.S. Bagnato,
  arXiv Prepr. p. 2005.11286 (2020)

\bibitem{Madeira2020c}
L.~Madeira, A.D. Garc{\'{i}}a-Orozco, F.E.A. dos Santos, V.S. Bagnato, Entropy
  \textbf{22}(9), 956 (2020)

\bibitem{Balantekin2019}
B.~Balantekin, C.~Bertulani, V.~Zelevinsky, Nucl. Phys. News \textbf{29}(3), 36
  (2019)

\end{thebibliography}

\end{document}